# Atomistic Processes of high-temperature plastic deformation of nanoscale body-centered cubic tungsten

Sixue Zheng[1], Zhengwu Fang[1], Scott X. Mao[1]*

[1]*Department of Mechanical Engineering and Materials Science, University of Pittsburgh,*

*Pittsburgh, Pennsylvania 15261, USA*

*Corresponding author: sxm2@pitt.edu;

# ABSTRACT

Much scientific and practical interest is currently focused on the atomic-scale mechanical behaviors of metallic nanocrystals with different crystal structures at room temperature, while the high-temperature plastic deformation in tungsten nanocrystals remains not well understood, due to the technical difficulty in elevating the experimental temperature during *in situ* mechanical tests in an extremely small chamber of transmission electron microscopes. In this study, a *in situ* high-temperature nanomechanical testing method is developed based on electrical-current-induced Joule heating in the metallic nanocrystal. By this method, it is found that three distinct deformation modes, that is deformation twinning, body-centered-cubic-face-centered-cubic-body-centered-cubic phase transformation and perfect dislocation slip, are sequentially activated in the tungsten nanocrystal during high-temperature tensile test. Such ductile behavior is related to not only the experimental temperature and but also the loading orientation. These findings shed light on the atomic-scale plastic deformation in body-centered cubic metals at elevated temperature.

## 1. Introduction

Many body-centered cubic (BCC) refractory metals, such as molybdenum (Mo) and tungsten (W), are critical constituents of high temperature alloys used in fusion reactors[1, 2] and turbine engines[3, 4]. BCC structured metal materials usually have high strength, Young's modulus and chemical durability, but they are brittle at room temperature and exhibit quite limited ductility at high temperature[232, 233, 234], which makes it difficult to investigate their ductile behaviors. In addition, the deformation characteristics of BCC metals are complex, due to the unique nonplanar nature of screw dislocation core[2, 5, 6, 7]. To date, compared to face-centered cubic (FCC) metals, the plastic deformation behaviors and the underlying mechanisms for BCC metals remain much less understood. Considering the widespread use of BCC metals in industrial applications, deep insights into their atomic-scale plastic deformation mechanisms are of vital importance.

Recent advances in *in situ* experimental mechanics open new avenues to investigate the mechanical behaviors of nanostructured metals[8, 9, 10, 11, 12, 13]. Some deformation behaviors and the underlying atomic-scale mechanisms in BCC nanostructured metals have been revealed, including surface dislocation nucleation[14], dislocation slip[11, 14, 15], deformation twinning[14, 16, 17], shear band dynamics[15, 18] and phase transformation[10, 19, 20, 21]. However, the prior studies focused on the mechanical properties and deformation behaviors of nanoscale BCC metals at room temperature. To date, the plastic behaviors of BCC nanocrystals at elevated temperature remain largely unexplored, especially at the atomic scale, due to the technical difficulties in achieving high temperature environmental conditions in an extremely small TEM chamber. It naturally raises the question of whether the plastic deformation mechanisms of BCC metals at room and elevated temperatures are different. Molecular dynamics (MD) simulations are usually employed to investigate the high-temperature deformation behaviors of nanoscale metal materials[22, 23, 24, 25].

Suffering from the inherent high strain rates, the accuracy of interatomic potentials[26, 27, 28] and the uncertainty in sample geometry[29], the computational results remain questionable, when they are directly extrapolated to those under the laboratory conditions. Thus, *in situ* atomic-scale high-temperature mechanical testing method is in urgent need for studying the high-temperature plastic deformation in BCC metallic nanocrystals to fill the knowledge gap.

Here, a novel testing method is developed, which allows *in situ* deformation at elevated temperature. This high-temperature testing method is based on the Joule heating effect produced by electrical current flow along the metallic nanocrystals. W nanocrystal is used as a model system to study the plastic deformation behaviors of BCC metallic nanocrystals. By conducting *in situ* tensile tests at elevated temperature and imaging using high resolution transmission electron microscopy (HRTEM), deformation twinning, BCC-FCC-BCC phase transformation and perfect dislocation slip are observed to sequentially occur in the W nanocrystals during high-temperature tensile test. Such ductile behavior is related to not only the experimental temperature and but also the loading orientation. Deep insights into the high-temperature ductile behaviors of BCC metals are of considerable importance from the standpoints of both scientific understanding and their practical applications.

## 2. Methods

The single-crystalline W nanocrystals were fabricated directly inside TEM using *in situ* nanowelding. Following Huang's testing method used to investigate the ductile behaviors of carbon nanotube at high temperature[240, 241, 242], *in situ* high-temperature nanomechanical testing method for metallic nanocrystals is developed here. The underlying concept of this method is that Joule heating, induced by the electrical current, could elevate the experimental temperature of the

W nanocrystal, and the temperature becomes steady in a quite short period of time[30] (Figure 1). In this study, Nanofactory TEM-STM (scanning tunneling microscope) holder allows for measuring the electrical current of the metallic nanocrystals under a constant bias voltage[31]. Then COMSOL Multiphysics based on finite element analysis (FEA) is utilized to provide accurate prediction of the temperature of the W nanocrystals, based on the thermal conductivity of the metallic nanocrystal (111.24 W·m$^{-1}$·K$^{-1}$ [32]), emissivity (0.35 [33]), boundary conditions, the measured electrical current and the bias voltage applied on the two ends of the sample[34]. Due to the fact that the sample in TEM chamber is under vacuum condition, natural convection can be ignored to simplify the simulation. The calculated temperatures of the 7.4-nm-diameter W nanocrystal (Figures 2 and Figure 4) during tensile test under the constant bias of 0.4 V and the electric current of 2.6 mA-3.7 mA are within the range from 1007 K to 1019 K, which are higher than the ductile-to-brittle transition temperature (DBTT) of nearly 500 K-700 K[35] for W. Thus, the bias voltage of 0.4 V is chosen in this study to investigate the high-temperature deformation behaviors of the 7.4-nm-diameter W nanocrystal. Likewise, the bias voltage of 0.35 V (the electric current of 3.5 mA-5.7 mA) is chosen for the 6.7-nm-diameter W nanocrystal (Figure 10). *In situ* tensile tests at high temperature are conducted inside a FEI Titan TEM equipped with a Nanofactory STM holder. The strain rate of 10$^{-3}$ s$^{-1}$ during tensile tests are controlled by the movement speed of the W nanoprobe. Moreover, all the tests are operated at 200 kV with low dose conditions (electron beam intensity < 10$^5$ A·m$^{-2}$) in a short time (less than 3 min) to minimize the potential beam effects on the deformation behavior of W nanocrystals[36, 37].

## 3. Results

### 3.1 Deformation twinning in W nanocrystal at high temperature

Figure 2 shows the atomic-scale deformation process of a 7.4-nm-diameter W nanocrystal at the strain rate of $10^{-3}$ s$^{-1}$ and under the temperature range of 1011 K to 1019 K. The loading direction is along [002] direction, and the viewing direction is along [1$\bar{1}$0] zone axis, which allows for observing perfect and partial dislocation slip during mechanical test. As shown in Figure 2a, the as-fabricated W nanocrystal is defect free. The gauge section, used to quantify the nanocrystal elongation, is selected based on the shape evolution of the W nanocrystal during tensile test[11, 13, 23]. During the entire loading process, plastic deformation occurs within the gauge section, and the diameters of the boundaries of the chosen gauge section remain nearly unchanged. Under tensile loading, the W nanocrystal firstly undergoes elastic deformation. As the lattice stress inside the W nanocrystal reaches the yielding point, a deformation twin with the operating twinning system of either (1$\bar{1}$2) [$\bar{1}$11] or ($\bar{1}$12) [1$\bar{1}$1] nucleates, resulting in a uniform elongation of 14.1% (Figure 2b and Figure 3a-b), which is determined by monitoring the change in gauge length. With the increase of tensile strain, the deformation twin grows along the loading direction, which is mediated by the consecutive slip of 1/6[$\bar{1}$11] or 1/6[1$\bar{1}$1] twinning partial dislocations (Figure 2c and Figure 2b-c). The completion of the twinning process ultimately results in the uniform elongation as large as 40% (Figure 2d and Figure 3d), which agrees well with the theoretical value of 41.4% resulted from twinning-induced lattice reorientation[38, 39]. Though the newly formed coherent twin boundaries, either (1$\bar{1}$2) or ($\bar{1}$12), are not in the edge-on condition when viewed along [1$\bar{1}$0] crystal direction, the occurrence of deformation twinning can be further confirmed through examining the crystallographic orientation relationship between the matrix and the twin. The enlarged HRTEM images (Figures 2e and 2g) and the corresponding fast Fourier transform pattern (Figures 2f and 2h) show that the axial direction of the W nanocrystal is reoriented from the original [002] to [$\bar{1}$10], and the viewing direction is changed from the original [1$\bar{1}$0] to the [001] zone axis, which is consistent with lattice reorientation caused by deformation twinning[22, 40]. Given that the Burgers vector directions of the twinning partials (1/6[$\bar{1}$11] and 1/6[1$\bar{1}$1]) are perpendicular to the direction of sample width ([110]), deformation twinning does not induce any change in sample diameter.

**3.2 BCC-FCC-BCC phase transformation in the W nanocrystal at high temperature**

After the completion of twinning process (Figure 4a), two regions (Region II and III) with different structural features appear in the twinned region (Region I) of W nanocrystal at the temperature of 1007 K-1011 K (Figure 4b), indicating the occurrence of phase transformation. A close observation of the crystal lattices in the different regions shows that the [001] lattices of BCC-structured W in Region I exhibit square-shape (the inset in Figure 4a), while the lattice shapes in Region II (the inset in Figure 4b) and Region III (the inset in Figure 4c) are rhombus. To identify the crystal structure of the new phases, the vectors between the transmission spot and the nearest (R1), second nearest (R2) and third nearest diffraction spots (R3) are examined in the fast Fourier transformed images of the three regions (Figure 4d-f). The ratios of R3/R1 and R2/R1 and the angle between R1 and R2 in Region II are 1.16, 1, 70.4°, respectively, which are different from the characteristic parameters of the <100>-BCC lattice (1, 1, 90° in Figure 4d), but match well with the characteristic parameters of the <110>-FCC lattice (1.15, 1, 70.5°)[41], suggesting that the crystal structure in Region II is FCC rather than BCC. Moreover, the experimental data for the lattice in Region III (1, 1, 60°) are consistent with the characteristic parameters of the <111>-BCC lattice (1, 1, 60°)[41]. Thus, the crystal lattice in Region III is determined to be BCC-structured W along <111> zone axis. In addition, Figure 4b and Figure 4d-f show that the orientation relation between the <001>-BCC lattice and the <110>-FCC lattice is consistent with the Nishiyama-Wassermann (N-W) relationship ([001]//[10$\bar{1}$] and (110)//(111))[42], and the orientation relationship between the <110>-FCC lattice and the <111>-BCC lattice follows the Kurdjumov-Sachs (K-S) relation ($[10\bar{1}]//[1\bar{1}1]$ and $(111)//(110)$)[43]. The good agreement with the well-established crystallographic relationship between FCC and BCC phases corroborates the above analysis on the crystal structures. The comparison between the filtered TEM images in the three regions (the insets in Figure 4a-c) and the simulated HRTEM images of FCC- and BCC-structured W (Figure 5) also demonstrates that the lattices in Region I, II and III should be <001>-BCC lattice, <110>-FCC lattice and <111>-BCC lattice, respectively. With further tensile loading, the <110>-FCC lattices, bridging the two BCC lattices, transform to new <111>-BCC lattices (Figure 4b-c), indicating that the FCC phase is a metastable intermediate phase for the phase transition from <001>-BCC lattice to <111>-BCC lattice, which is similar to the two-step BCC-FCC-BCC phase transition observed in Nb nanowire[19] and Mo thin film[21]. The first phase transition from <001>-BCC lattice to <110>-FCC lattice follows the N-W transformation process, and the second phase transition from <110>-FCC lattice to <111>-BCC lattice follows the K-S transformation process.

In addition to the BCC-FCC-BCC phase transformation, perfect dislocation slip occurs in Region III with <111>-BCC lattice, as demonstrated by the observation of a perfect dislocation (the inset in Figure 4c) and the increase in the height of a surface step (Figure 4b-c) [18, 19]. After tensile failure, the BCC-structured W along <111> zone axis could stably exist in the fractured nanocrystal under zero external stress at room temperature (Figure 6a). The fast Fourier transform patterns of the crystal lattice in the fracture nanocrystal (Figure 6b) is in good agreement with the transmission electron pattern of BCC-structured W in [111] zone axis[41]. To further validate the above analysis on the crystalline structure in Region III, the distances between the adjacent atomic columns along <112> directions are carefully measured (Figure 6c-e), based on the difference in the contrast of the areas inside and outside the atomic columns[11, 13, 23]. To minimize the measurement errors, the lattice spacing measurements for 5 atomic columns are conducted in this study[44, 45] (Figure 6c-e). The lattice distances of the two neighboring atomic columns along <112> directions (indicated by the yellow lines in Figure 6a) are determined to be 2.6 Å, and the interplanar angles between the {110} planes are measured to be 60°, which are consistent with the characteristic parameters of the <111>-BCC lattice (2.589 Å and 60° in Figure 5f). It further proves the occurrence of BCC-FCC-BCC phase transformation in W nanocrystal.

Based on the above microstructural analysis, it is demonstrated that deformation twinning, BCC-FCC-BCC phase transformation and perfect dislocation slip sequentially occur in the 7.4-nm-diameter W nanocrystal during high-temperature tensile test (Figure 2 and Figure 4). Different from the high-temperature ductile behaviors, only deformation twinning is observed to occur in the W nanocrystal under [002] tensile loading at room temperature, as shown in Figure 7. Moreover, the room-temperature tensile test of the <110>-oriented W nanocrystal shows that perfect dislocation slip on {110} planes is the dominant mechanism, and no phase transformation is observed (Figure 8). Thus, the experimental observation of BCC-FCC-BCC phase transformation in W nanocrystal is related to the experimental temperature.

**3.3 Orientation dependence of the deformation behaviors of W nanocrystals**

To probe the orientation dependence of the deformation behaviors of W nanocrystals, *in situ* tensile test of a [112]-oriented W nanocrystal with the diameter of 8.7 nm is conducted at room temperature under a strain rate of $10^{-3}$ s$^{-1}$ (Figure 9). The fast Fourier transform pattern of the pristine W nanocrystal shows that the axial and viewing directions of the nanocrystal are parallel to [112] and [$\bar{1}\bar{1}$1] direction, respectively (Figure 9a). Before tensile test, no lattice defects are observed in the as-fabricated W nanocrystal. Upon the yielding of the W nanocrystal, 1/2[$\bar{1}$11]- type dislocations nucleate from the free surface and then glide on (101) slip planes, as shown in Figure 9b. The annihilation of these dislocations at the opposite free surface results in the formation and enlargement of the surface step (Figure 9b). Further tensile deformation of the W nanocrystal is mediated by perfect dislocation slip, causing continuous decrease in the nanocrystal width (Figure 9c-f). The room-temperature deformation behaviors of W nanocrystals are consistent with the previous experimental observation of dislocation-dominated plasticity in BCC nanocrystals[14, 15, 18]. The fast Fourier transform pattern of the deformed W nanocrystal (the insets in Figure 9f) shows that no phase transformation occurs in the W nanocrystal during tensile loading.

To investigate the high-temperature deformation behaviors of W nanocrystals under [112] tension, the tensile test of a 6.7-nm-diameter W nanocrystal is performed at the strain rate of $10^{-3}$ $s^{-1}$ under the temperature range of 912 K to 915 K (Figure 10). Similar to the room-temperature deformation behaviors, plastic deformation in the W nanocrystal at high temperature is mediated by the slip of 1/2<111> perfect dislocations, including the nucleation of dislocations from free surface, the subsequent dislocation propagation across the nanocrystal and dislocation annihilation at free surface (Figure 10). The fast Fourier transform patterns of the pristine and deformed W nanocrystals (the insets in Figure 10a and 10f) show that the lattice structures of the W nanocrystals remain unchanged during the high-temperature tensile test. Thus, no phase transformation occurs in the [112]-oriented W nanocrystals at room and high temperatures. The BCC-FCC-BCC phase transformation could only occur in the W nanocrystal under [110] tensile loading at high temperature. Thus, the occurrence of BCC-FCC-BCC phase transformation in W nanocrystal is related to the loading direction and the experimental temperature. It should be noted that dislocation starvation is not always observed in the W nanocrystals during mechanical loading

(Figure 8, Figure 9 and Figure 10), which is associated with the low mobility of screw dislocations with non-planar core structure[46, 47, 48, 49, 50].

## 4. Discussion

To date, a wealth of experimental and theoretical studies on the deformation behaviors of bulk BCC metals[51, 52, 53, 54, 55, 56] and BCC nanopillars[46, 47, 48, 57, 58, 59] showed that 1/2<1 1 1> screw dislocation was the only plastic deformation carrier. As the microstructural feature size of BCC metal decreases to nanoscale, the high stress, resulting from the well-established 'smaller is stronger' norm, facilitates the activation of deformation twinning and phase transformation[14, 16, 17, 19, 21, 60]. The experimental observation of BCC-FCC-BCC phase transformation in this study is also attributed to the high lattice stress accumulated in the twinned region of the W nanocrystal during tensile deformation. The high stress in the W nanocrystal could provide sufficient energy to stimulate lattice instabilities and serves as a thermodynamic driving force for overcoming the energy barrier for phase transformation[19, 21, 60]. Furthermore, according to transition state theory, the increase in temperature and stress could lower the activation energy for structural transformation[61, 62, 63, 64, 65]. Thus, phase transformation is energetically favorable in the deformed W nanocrystal at high temperature. In addition to the high stress and the experimental temperature, the loading orientation is also a critical factor influencing the activation of phase transformation. In this study, the applied tensile loading direction for the W nanocrystal is parallel to the BCC (110) and the FCC (111) planes (Figure 4). Given that the yield strength for W nanocrystal is ~19.2 GPa[14], the extremely high resolved shear stress (~19.2 GPa) on the BCC (110) and the FCC (111) planes could induce successive atomic shear on the (110) and the (111) planes, resulting in the occurrence of BCC-FCC-BCC phase transformation[21, 66].

This work reveals that the loading orientation significantly influences the dominant deformation mode in nanoscale W. Different from FCC metals[67, 68], the orientation-dependent deformation behaviors observed in W nanocrystal cannot be well explained by Schmid factor analysis, due to the interplay between the non-planar screw dislocation core and the applied stress[17, 69]. In BCC metals, the dominant deformation mechanism is related to the so-called twinning-antitwinning slip asymmetry of 1/6<111> partial dislocations on {112} planes[70, 71, 72, 73]. Under uniaxial tensile loading, the <100>-oriented W nanocrystal is in the twinning-orientation favoring the activation of deformation twinning, while the <110>- and <112>-oriented W nanocrystals are in the antitwinning-orientations favoring the activation of perfect dislocation slip. Furthermore, the largest Schmid factors for dislocation slip and twinning in <100>-oriented W nanocrystal are 0.41 and 0.47, respectively, further facilitating the operation of deformation twinning. In contrast, the largest Schmid factors for slip and twinning in <112>-oriented W nanocrystal are 0.41 and 0.39, respectively, facilitating the activation of perfect dislocation slip. As a result, W nanocrystal with <100> orientation is observed to deform by deformation twinning on {112} planes (Figure 2 and Figure 4), whereas the deformation in <110>- and <112>-oriented W nanocrystal is mediated by perfect dislocation slip on {110} planes (Figure 8, Figure 9 and Figure 10), which is consistent with the previous experimental and computational results[12, 14, 25, 50, 70, 74, 75, 76].

In addition to the loading orientation, experimental temperature is another critical factor influencing the deformation behaviors of W nanocrystals. It has been well documented that the tendency for twinning in bulk FCC and BCC metals usually decreases with increasing experimental temperature[77, 78]. However, previous MD simulations revealed that deformation twinning occurred in W nanowire at temperature higher than 1500 K[22]. Consistent with the previous computational results, this study provides the direct experimental evidence that deformation twinning is the dominant deformation mechanism in W nanocrystals at high temperature (1007-1019 K). After the completion of twinning process, BCC-FCC-BCC phase transformation subsequently occurs in the twinned region to mediate the plasticity of W nanocrystal. Moreover, computational studies on the tensile behavior of W nanowire revealed that the [112]-oriented W nanowire failed in a brittle manner at 300 K but deformed by twinning on {112} planes at 800 K, exhibiting the temperature-dependent brittle-to-ductile transition in deformation behaviors[79]. Different from the computational results, the experimental observations in this study show that perfect dislocation slip mediates the plasticity of the [112]-oriented W nanocrystals at both room (300 K) and high temperatures (912 K-915 K). Such difference between the experimental and computational results is probably attributed to the high strain rates used in MD simulations[8, 11, 63, 80] and the uncertainty in interatomic potentials[26, 27, 28] and sample geometry[29]. Directly validating the research findings from computational investigations, the *in situ* high-temperature nanomechanical testing method developed in this study opens new avenues to investigate the high-temperature deformation behaviors of nanostructured metals at the atomic scale.

## 5. Conclusion

In conclusion, a *in situ* high-temperature nanomechanical testing method is proposed, which is based on the Joule heating effect caused by the electrical current through W nanocrystals. Using such method, the atomic-scale deformation behaviors of W nanocrystals at high temperature are investigated. Deformation twinning, BCC–FCC-BCC transformation and perfect dislocation slip are sequentially activated to mediate the plasticity of W nanocrystal. The dominant deformation mode in W nanocrystals is related to the loading orientation and the experimental temperature. This *in situ* study sheds light on understanding the high-temperature deformation behaviors of BCC metals at the atomic scale.

**Author Contributions**

S.Z. conducted the *in situ* TEM experiments and performed the experimental data analysis. S.Z., Z.F. and S.X.M. prepared the paper with the contribution of all authors.

**Acknowledgements**

S.X.M. acknowledges support from the NSF 1760916 through University of Pittsburgh.

**Conflict of interest**

The authors declare no competing financial interest.

**Reference**

1. Zinkle S, Ghoniem N. Operating temperature windows for fusion reactor structural materials. *Fusion Engineering and design* 2000, **51:** 55-71.

2. Xu K, Niu L-L, Jin S, Shu X, Xie H, Wang L*, et al.* Atomistic simulations of screw dislocations in bcc tungsten: From core structures and static properties to interaction with vacancies. *Nuclear Instruments and Methods in Physics Research Section B: Beam Interactions with Materials and Atoms* 2017, **393:** 174-179.

3. El-Genk MS, Tournier J-M. A review of refractory metal alloys and mechanically alloyed-oxide dispersion strengthened steels for space nuclear power systems. *Journal of Nuclear materials* 2005, **340**(1)**:** 93-112.

4. Perepezko JH. The hotter the engine, the better. *Science* 2009, **326**(5956)**:** 1068-1069.

5. Rao S, Woodward C. Atomistic simulations of (a/2) 〈111〉 screw dislocations in bcc Mo using a modified generalized pseudopotential theory potential. *Philosophical Magazine A* 2001, **81**(5)**:** 1317-1327.

6. Yang L, Söderlind P, Moriarty JA. Accurate atomistic simulation of (a/2)〈111〉 screw dislocations and other defects in bcc tantalum. *Philosophical Magazine A* 2001, **81**(5)**:** 1355-1385.

7. Vitek V. Core structure of screw dislocations in body-centred cubic metals: relation to symmetry and interatomic bonding. *Philos Mag* 2004, **84**(3-5)**:** 415-428.

8. Zheng S, Mao SX. Advances in experimental mechanics at atomic scale. *Extreme Mechanics Letters* 2021, **45:** 101284.

9. Wang X, Zheng S, Deng C, Weinberger CR, Wang G, Mao SX. In Situ Atomic-Scale Observation of 5-Fold Twin Formation in Nanoscale Crystal under Mechanical Loading. *Nano Letters* 2023.

10. Zheng S, Wang X, Tan S, Wang G, Mao SX. Atomistic processes of diffusion-induced unusual compression fracture in metallic nanocrystals. *Mater Res Lett* 2022, **10**(12)**:** 805-812.

11. Zheng S, Shinzato S, Ogata S, Mao S. Experimental molecular dynamics for individual atomic-scale plastic events in nanoscale crystals. *Journal of the Mechanics and Physics of Solids* 2022, **158:** 104687.

12. Zheng S, Mao SX. In situ atomic-scale observation of dislocation-mediated discrete plasticity in nanoscale crystals. *Material Science & Engineering International Journal* 2021, **5**(3)**:** 93-94.

13. Wang X, Zheng S, Shinzato S, Fang Z, He Y, Zhong L*, et al.* Atomistic processes of surface-diffusion-induced abnormal softening in nanoscale metallic crystals. *Nature Communications* 2021, **12**(1)**:** 1-9.

14. Wang J, Zeng Z, Weinberger CR, Zhang Z, Zhu T, Mao SX. In situ atomic-scale observation of twinning-dominated deformation in nanoscale body-centred cubic tungsten. *Nature Materials* 2015, **14**(6)**:** 594-600.

15. Wei S, Wang Q, Wei H, Wang J. Bending-induced deformation twinning in body-centered cubic tungsten nanowires. *Mater Res Lett* 2019, **7**(5)**:** 210-216.

16. Wang J, Faisal AH, Li X, Hong Y, Zhu Q, Bei H*, et al.* Discrete twinning dynamics and size-dependent dislocation-to twin transition in body-centred cubic tungsten. *Journal of Materials Science & Technology* 2022, **106:** 33-40.

17. Wang J, Zeng Z, Wen M, Wang Q, Chen D, Zhang Y*, et al.* Anti-twinning in nanoscale tungsten. *Science Advances* 2020, **6**(23)**:** eaay2792.

18. Wang J, Wang Y, Cai W, Li J, Zhang Z, Mao SX. Discrete shear band plasticity through dislocation activities in body-centered cubic tungsten nanowires. *Scientific Reports* 2018, **8**(1)**:** 1-8.

19. Wang Q, Wang J, Li J, Zhang Z, Mao SX. Consecutive crystallographic reorientations and superplasticity in body-centered cubic niobium nanowires. *Science Advances* 2018, **4**(7)**:** eaas8850.

20. Zhang J, Li Y, Li X, Zhai Y, Zhang Q, Ma D, *et al.* Timely and atomic-resolved high-temperature mechanical investigation of ductile fracture and atomistic mechanisms of tungsten. *Nature Communications* 2021, **12**(1)**:** 1-10.

21. Wang SJ, Wang H, Du K, Zhang W, Sui ML, Mao SX. Deformation-induced structural transition in body-centred cubic molybdenum. *Nature Communications* 2014, **5**(1)**:** 1-9.

22. Li SZ, Ding XD, Deng JK, Lookman T, Li J, Ren XB, *et al.* Superelasticity in bcc nanowires by a reversible twinning mechanism. *Phys Rev B* 2010, **82**(20)**:** 205435.

23. Zhong L, Sansoz F, He Y, Wang C, Zhang Z, Mao SX. Slip-activated surface creep with room-temperature super-elongation in metallic nanocrystals. *Nat Mater* 2017, **16**(4)**:** 439-445.

24. Sun J, He L, Lo Y-C, Xu T, Bi H, Sun L, *et al.* Liquid-like pseudoelasticity of sub-10-nm crystalline silver particles. *Nature materials* 2014, **13**(11)**:** 1007-1012.

25. Feng Y-X, Shang J-X, Qin S-J, Lu G-H, Chen Y. Twin and dislocation mechanisms in tensile W single crystal with temperature change: a molecular dynamics study. *Physical Chemistry Chemical Physics* 2018, **20**(26)**:** 17727-17738.

26. Deng C, Sansoz F. Fundamental differences in the plasticity of periodically twinned nanowires in Au, Ag, Al, Cu, Pb and Ni. *Acta Mater* 2009, **57**(20)**:** 6090-6101.

27. Zheng H, Cao A, Weinberger CR, Huang JY, Du K, Wang J, *et al.* Discrete plasticity in sub-10-nm-sized gold crystals. *Nature Communications* 2010, **1**.

28. Park N-Y, Nam H-S, Cha P-R, Lee S-C. Size-dependent transition of the deformation behavior of Au nanowires. *Nano Research* 2015, **8**(3)**:** 941-947.

29. Wang J, Sansoz F, Huang J, Liu Y, Sun S, Zhang Z, *et al.* Near-ideal theoretical strength in gold nanowires containing angstrom scale twins. *Nat Commun* 2013, **4:** 1742.

30. Oshima Y, Kurui Y. In situ TEM observation of controlled gold contact failure under electric bias. *Phys Rev B* 2013, **87**(8)**:** 081404.

31. Zhao J, Sun H, Dai S, Wang Y, Zhu J. Electrical breakdown of nanowires. *Nano letters* 2011, **11**(11)**:** 4647-4651.

32. Stojanovic N, Maithripala D, Berg J, Holtz M. Thermal conductivity in metallic nanostructures at high temperature: Electrons, phonons, and the Wiedemann-Franz law. *Phys Rev B* 2010, **82**(7)**:** 075418.

33. Ahmad I, Khalid S, Khawaja E. Filament temperature of low power incandesecent lamps: Stefan-Boltzmann law. 2010.

34. Chen H, Luo X, Wang D, Ziegler M, Huebner U, Zhang B*, et al.* Length-scale dominated thermal fatigue behavior in nanocrystalline Au interconnect lines. *Materialia* 2019, **7:** 100337.

35. Gumbsch P, Riedle J, Hartmaier A, Fischmeister HF. Controlling factors for the brittle-to-ductile transition in tungsten single crystals. *Science* 1998, **282**(5392)**:** 1293-1295.

36. Sun J, He L, Lo Y-C, Xu T, Bi H, Sun L*, et al.* Liquid-like pseudoelasticity of sub-10-nm crystalline silver particles. *Nature materials* 2014, **13**(11)**:** 1007.

37. Zhong L, Sansoz F, He Y, Wang C, Zhang Z, Mao SX. Slip-activated surface creep with room-temperature super-elongation in metallic nanocrystals. *Nature materials* 2017, **16**(4)**:** 439-445.

38. Wang P, Chou W, Nie A, Huang Y, Yao H, Wang H. Molecular dynamics simulation on deformation mechanisms in body-centered-cubic molybdenum nanowires. *J Appl Phys* 2011, **110**(9)**:** 093521.

39. Cao A. Shape memory effects and pseudoelasticity in bcc metallic nanowires. *J Appl Phys* 2010, **108**(11)**:** 113531.

40. Wang P, Chou W, Nie AM, Huang Y, Yao HM, Wang HT. Molecular dynamics simulation on deformation mechanisms in body-centered-cubic molybdenum nanowires. *J Appl Phys* 2011, **110**(9)**:** 093521.

41. Williams DB, Carter CB. *Transmission Electron Microscopy*, 2 edn. Springer New York: New York, 2009.

42. Nishiyama Z. X-ray investigation of the mechanism of the transformation from face centered cubic lattice to body centered cubic. *Sci Rep Tohoku Univ* 1934, **23:** 637-664.

43. Kurdjumov G, Sachs G. Over the mechanisms of steel hardening. *Z Phys* 1930, **64:** 325-343.

44. Wang J, Sansoz F, Huang J, Liu Y, Sun S, Zhang Z*, et al.* Near-ideal theoretical strength in gold nanowires containing angstrom scale twins. *Nature Communications* 2013, **4**(1)**:** 1-8.

45. Wang J, Sansoz F, Deng C, Xu G, Han G, Mao SX. Strong Hall-Petch Type Behavior in the Elastic Strain Limit of Nanotwinned Gold Nanowires. *Nano Letters* 2015, **15**(6)**:** 3865-3870.

46. Greer JR, Weinberger CR, Cai W. Comparing the strength of fcc and bcc sub-micrometer pillars: Compression experiments and dislocation dynamics simulations. *Materials Science and Engineering: A* 2008, **493**(1-2)**:** 21-25.

47. Weinberger CR, Cai W. Surface-controlled dislocation multiplication in metal micropillars. *Proc Natl Acad Sci U S A* 2008, **105**(38)**:** 14304-14307.

48. Kim JY, Jong DC, Greer JR. Tensile and compressive behavior of tungsten, molybdenum, tantalum and niobium at the nanoscale. *Acta Mater* 2010, **58**(7)**:** 2355-2363.

49. Huang L, Li QJ, Shan ZW, Li J, Sun J, Ma E. A new regime for mechanical annealing and strong sample-size strengthening in body centred cubic molybdenum. *Nat Commun* 2011, **2**(1)**:** 1-6.

50. Sainath G, Choudhary B. Molecular dynamics simulations on size dependent tensile deformation behaviour of [110] oriented body centred cubic iron nanowires. *Materials Science and Engineering: A* 2015, **640:** 98-105.

51. Caillard D. A TEM in situ study of alloying effects in iron. II—Solid solution hardening caused by high concentrations of Si and Cr. *Acta Mater* 2013, **61**(8)**:** 2808-2827.

52. Couzinié J-P, Lilensten L, Champion Y, Dirras G, Perrière L, Guillot I. On the room temperature deformation mechanisms of a TiZrHfNbTa refractory high-entropy alloy. *Materials Science and Engineering: A* 2015, **645:** 255-263.

53. Lei Z, Liu X, Wu Y, Wang H, Jiang S, Wang S*, et al.* Enhanced strength and ductility in a high-entropy alloy via ordered oxygen complexes. *Nature* 2018, **563**(7732)**:** 546-550.

54. Lilensten L, Couzinie J-P, Perriere L, Hocini A, Keller C, Dirras G*, et al.* Study of a bcc multi-principal element alloy: Tensile and simple shear properties and underlying deformation mechanisms. *Acta Mater* 2018, **142:** 131-141.

55. Gröger R, Vitek V. Multiscale modeling of plastic deformation of molybdenum and tungsten. III. Effects of temperature and plastic strain rate. *Acta Mater* 2008, **56**(19)**:** 5426-5439.

56. Caillard D. Kinetics of dislocations in pure Fe. Part I. In situ straining experiments at room temperature. *Acta Mater* 2010, **58**(9)**:** 3493-3503.

57. Kim J-Y, Jang D, Greer JR. Insight into the deformation behavior of niobium single crystals under uniaxial compression and tension at the nanoscale. *Scripta Mater* 2009, **61**(3)**:** 300-303.

58. Schneider A, Kaufmann D, Clark B, Frick C, Gruber P, Mönig R, *et al.* Correlation between critical temperature and strength of small-scale bcc pillars. *Physical review letters* 2009, **103**(10)**:** 105501.

59. Marichal C, Van Swygenhoven H, Van Petegem S, Borca C. {110} Slip with {112} slip traces in bcc Tungsten. *Scientific reports* 2013, **3**(1)**:** 1-7.

60. Lu Y, Sun S, Zeng Y, Deng Q, Chen Y, Li Y*, et al.* Atomistic mechanism of nucleation and growth of a face-centered orthogonal phase in small-sized single-crystalline Mo. *Mater Res Lett* 2020, **8**(9)**:** 348-355.

61. Chen LY, He MR, Shin J, Richter G, Gianola DS. Measuring surface dislocation nucleation in defect-scarce nanostructures. *Nat Mater* 2015, **14**(7)**:** 707-713.

62. Zhu T, Li J, Samanta A, Leach A, Gall K. Temperature and strain-rate dependence of surface dislocation nucleation. *Physical Review Letters* 2008, **100**(2)**:** 025502.

63. Zhu T, Li J. Ultra-strength materials. *Progress in Materials Science* 2010, **55**(7)**:** 710-757.

64. Li J, Sarkar S, Cox WT, Lenosky TJ, Bitzek E, Wang Y. Diffusive molecular dynamics and its application to nanoindentation and sintering. *Phys Rev B* 2011, **84**(5)**:** 054103.

65. Li J. The mechanics and physics of defect nucleation. *MRS bulletin* 2007, **32**(2)**:** 151-159.

66. Weidner A. *Deformation Processes in TRIP/TWIP Steels: In-situ characterization techniques*, 1 edn. Springer Cham, 2020.

67. Zheng S, Luo X, Wang D, Zhang G. A novel evaluation strategy for fatigue reliability of flexible nanoscale films. *Mater Res Express* 2018, **5**(3)**:** 035012.

68. Zheng S, Luo X, Zhang G. Cumulative shear strain–induced preferential orientation during abnormal grain growth near fatigue crack tips of nanocrystalline Au films. *J Mater Res* 2020, **35**(4)**:** 1-8.

69. Gröger R, Vitek V. Breakdown of the Schmid law in bcc molybdenum related to the effect of shear stress perpendicular to the slip direction.  Materials Science Forum; 2005: Trans Tech Publ; 2005. p. 123-126.

70. Sainath G, Choudhary BK. Orientation dependent deformation behaviour of BCC iron nanowires. *Computational Materials Science* 2016, **111:** 406-415.

71. Narayanan S, McDowell DL, Zhu T. Crystal plasticity model for BCC iron atomistically informed by kinetics of correlated kinkpair nucleation on screw dislocation. *Journal of the Mechanics and Physics of Solids* 2014, **65:** 54-68.

72. Duesbery MS, Vitek V. Plastic anisotropy in bcc transition metals. *Acta Mater* 1998, **46**(5)**:** 1481-1492.

73. Ito K, Vitek V. Atomistic study of non-Schmid effects in the plastic yielding of bcc metals. *Philosophical Magazine A* 2001, **81**(5)**:** 1387-1407.

74. Healy CJ, Ackland GJ. Molecular dynamics simulations of compression–tension asymmetry in plasticity of Fe nanopillars. *Acta Mater* 2014, **70:** 105-112.

75. Sainath G, Choudhary B, Jayakumar T. Molecular dynamics simulation studies on the size dependent tensile deformation and fracture behaviour of body centred cubic iron nanowires. *Computational Materials Science* 2015, **104:** 76-83.

76. Sainath G, Choudhary B. Atomistic simulations on ductile-brittle transition in⟨ 111⟩ BCC Fe nanowires. *J Appl Phys* 2017, **122**(9)**:** 095101.

77. Zhu Y, Liao X, Wu X. Deformation twinning in nanocrystalline materials. *Progress in Materials Science* 2012, **57**(1)**:** 1-62.

78. Christian JW, Mahajan S. Deformation twinning. *Progress in materials science* 1995, **39**(1-2)**:** 1-157.

79. Xu S, Su Y, Chen D, Li L. An atomistic study of the deformation behavior of tungsten nanowires. *Applied Physics A* 2017, **123**(12)**:** 1-9.

80. Wang J, Mao SX. Atomistic perspective on in situ nanomechanics. *Extreme Mechanics Letters* 2016, **8:** 127-139.

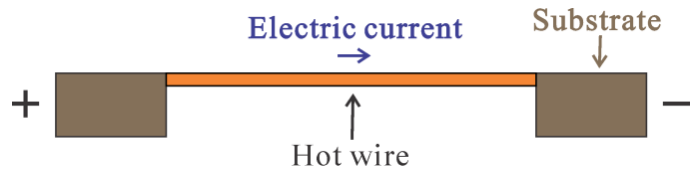

**Figure 1.** The experimental temperature is elavated due to the Joule heating effect induced by the electrical current through the metallic nanocrystal.

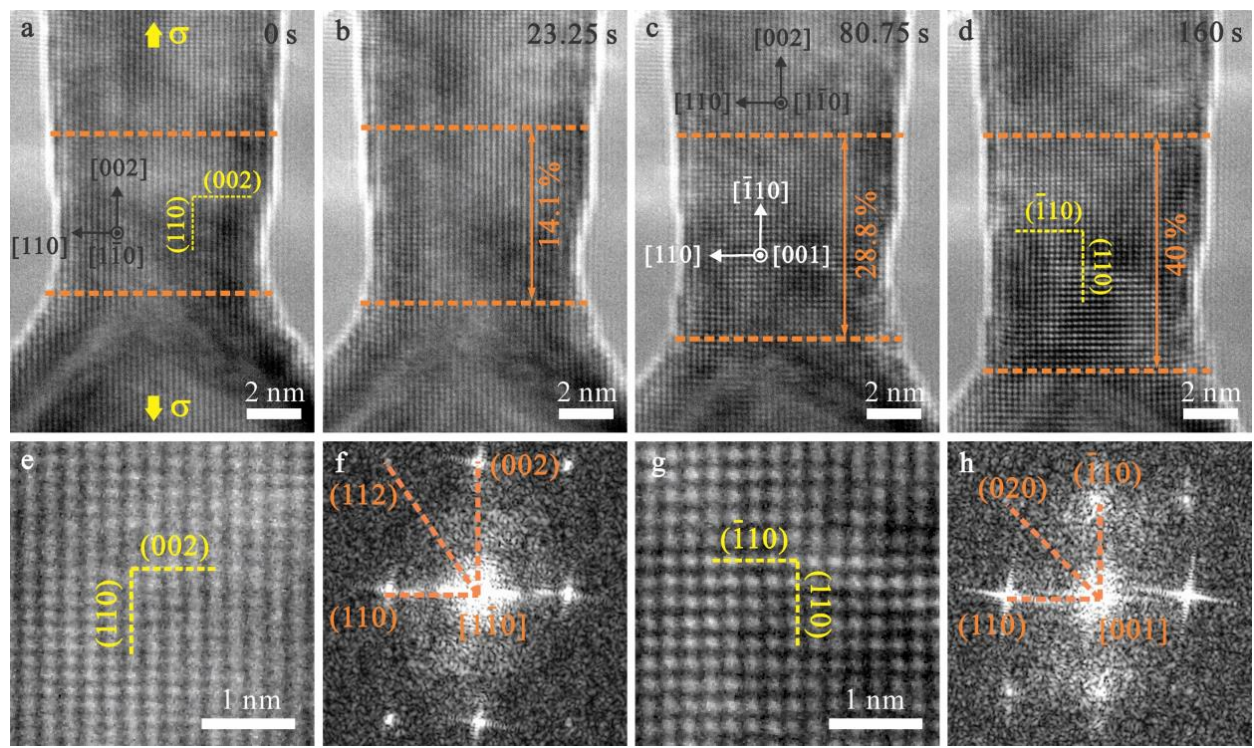

**Figure 2. Deformation twinning in the 7.4-nm-diameter W nanocrystal under [002] tensile loading at the strain rate of $10^{-3}$ s$^{-1}$ and under the temperature of 1011 K-1019 K.** (a) High-resolution TEM image of the pristine W nanocrystal. The viewing and loading directions are along [1$\bar{1}$0] and [002], rexpectively. The dashed lines indicate the gauge section used for calculating the elongation strain during tensile test. (b) Nucleation of a deformation twin with a twinning system of (11$\bar{2}$) [1$\bar{1}$1] or (1$\bar{1}$2) [11$\bar{1}$], causing a sudden elongation of 14.1%. (c) Growth of the deformation twin during further tensile loading. (d) The deformed W nanocrystal after deformation twinning exhibites an uniform elongation of nearly 40%. (e-f) Magnified high resolution TEM image (e) and fast Fourier transform pattern (f) of the pristine W nanocrystal before tensile test. (g-h) Enlarged high resolution TEM image (g) and fast Fourier transform pattern (h) of the area in the deformation twin.

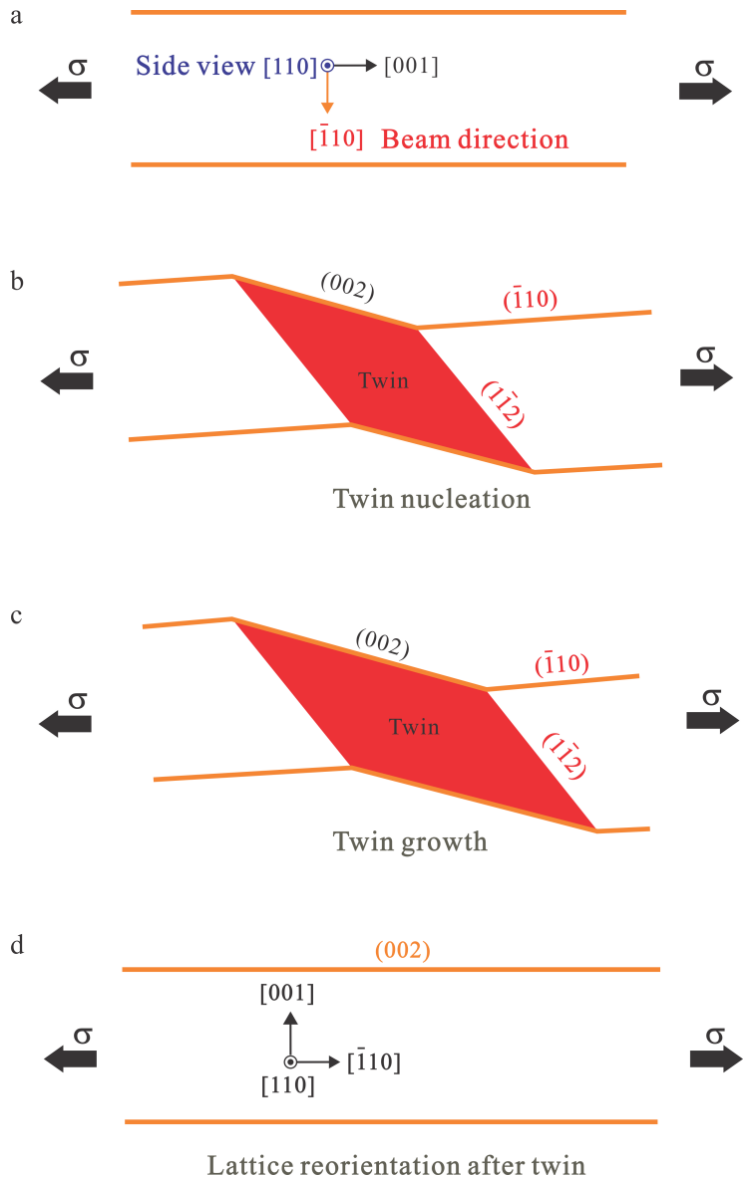

**Figure 3.** Schematics showing the process of deformation twinning in W nanocrystal, when viewed along the [110] zone axis. One of the two possible equivalent twinning systems, namely $1/6[\bar{1}11](1\bar{1}2)$, is selected.

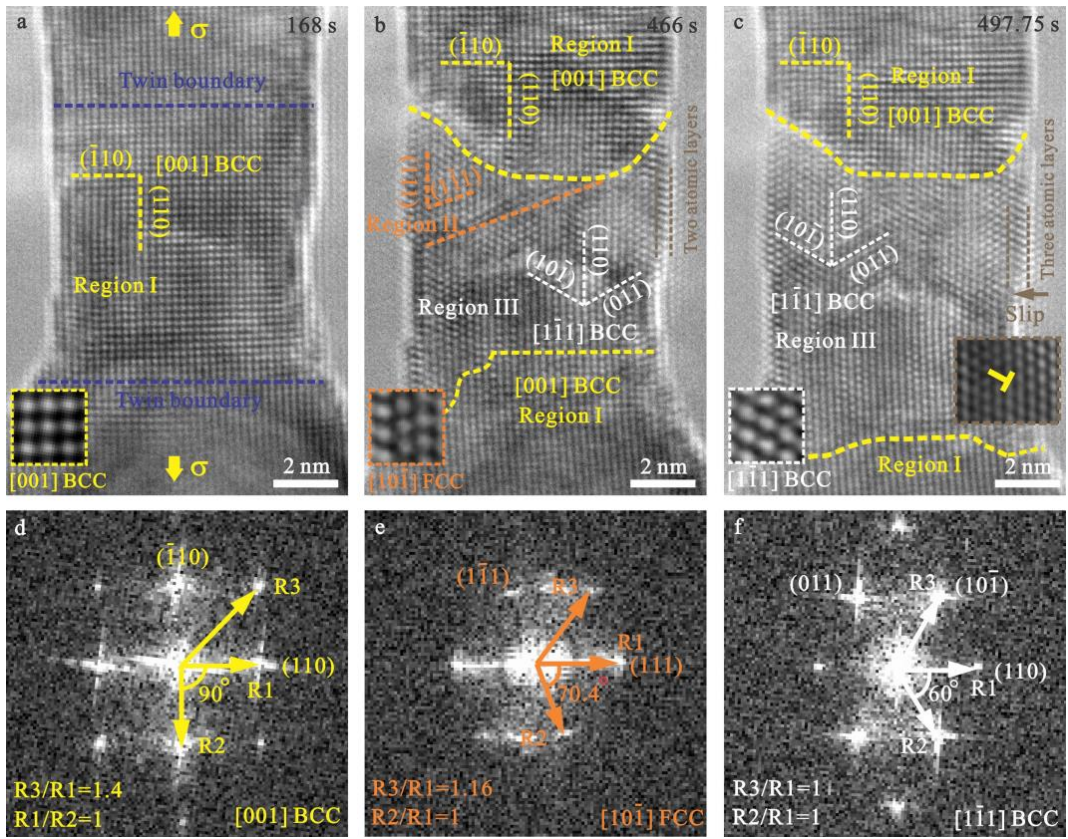

**Figure 4.** BCC-FCC-BCC phase transformation in 7.4-nm-diameter W nanocrystal within the temperature range from 1007 K to 1011 K. (a) TEM image of the deformed W nanocrystal after deformation twinning. Filtered TEM image in the inset shows the atomic structures of the <001>-BCC lattice. (b-c) Occurrence of BCC-FCC-BCC phase transition. Filtered TEM images in the insets of (b) and (c) show the atomic structures of the <110>-FCC lattices and <111>-BCC lattices, respectively. The red dashed and yellow dashed lines represent the <111>-BCC/<110>-FCC and the <110>-FCC/<111>-BCC interfaces, respectively. After phase transformation, the loading direction is reoriented from <110> to <112>, and the viewing direction changes from <001> to <111>. Perfect dislocation slip is observed in <111>-BCC lattice, causing the increase in the surface step hight by one atomic layer. The TEM image in the inset of (c) shows the surface nucleated perfect dislocation. (d-f) Fast Fourier transform patterns of the <001>-BCC, <110>-FCC and <111>-BCC lattices.

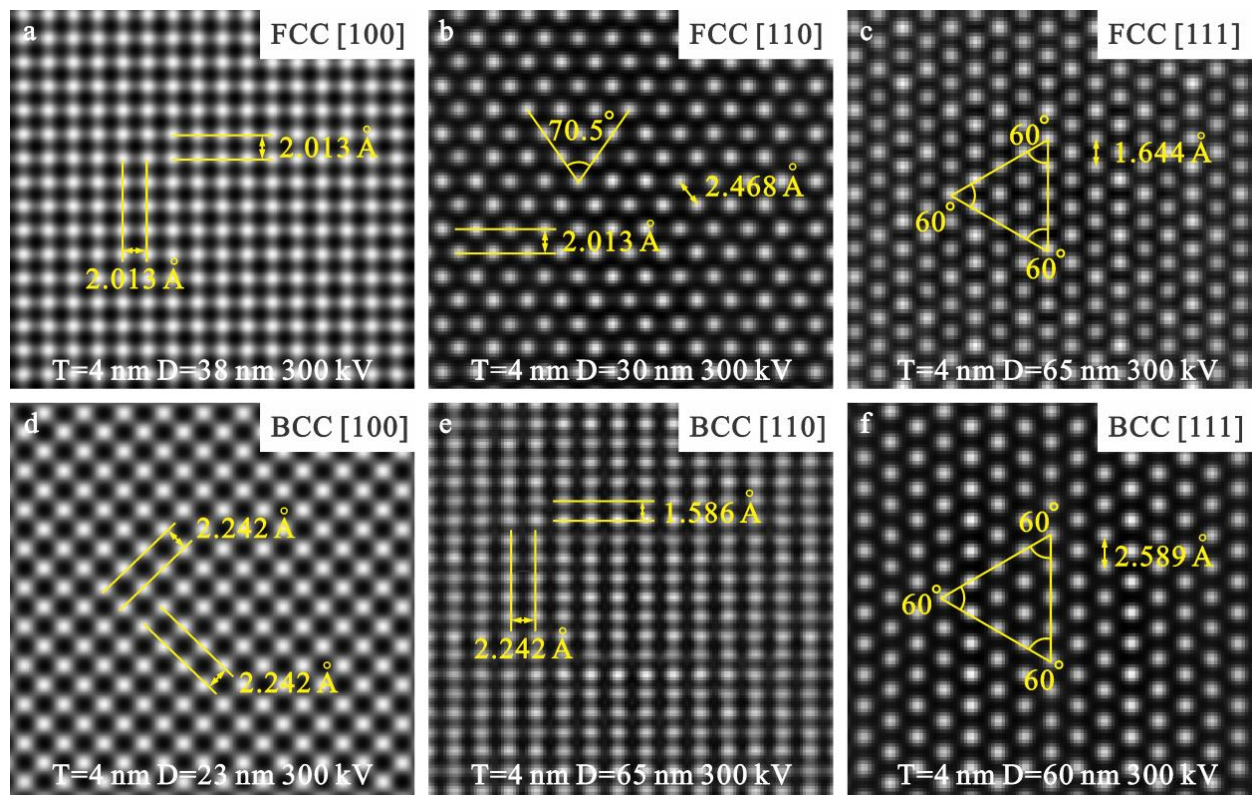

**Figure 5.** Simulated HRTEM images of FCC- and BCC-structured W viewed along several low-indexed zone axes. The interplanar angles and the projecting lattice distances of the adjacent atomic coloumns along different directions are both indicated. By comparing the experimental observation of the new phases in the W nanocrystal with the simulated HRTEM images, the new phases can be determined to be FCC-structured W in the <110> zone axis and BCC-structured W in the <111> zone axis. "T" and "D" in all the images indicate the thickness of the simulated W nanocrystal and the defocus distance used in simulation, respectively.

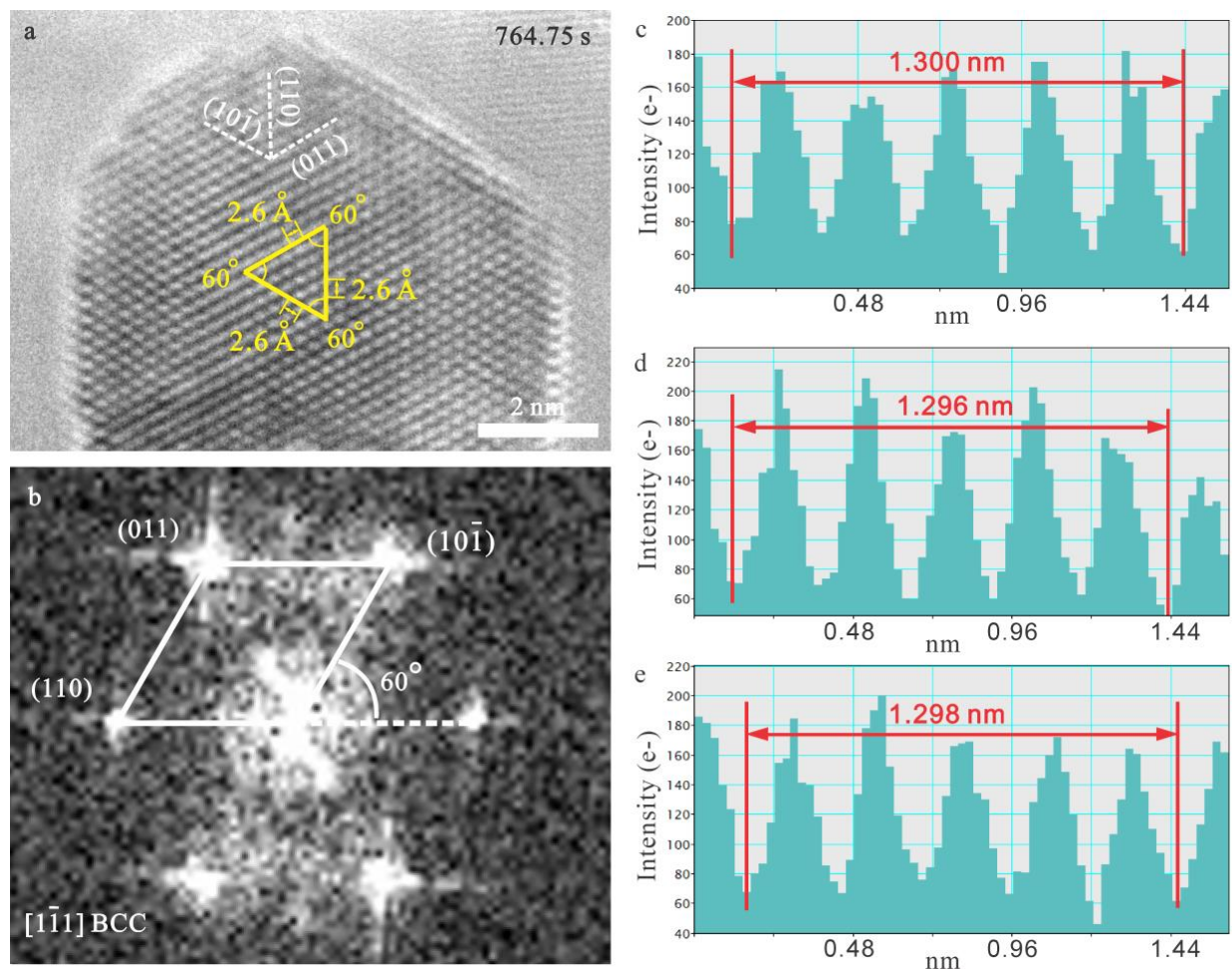

**Figure 6. <111>-BCC phase in the fractured W nanocrystal.** (a) TEM image of the fractured W nanocrystal showing that <111>-BCC phase could exist for a long time under zero external stress at room temperature. (b) Fast Fourier transform patterns of (a), which is in good agreement with the transmission electron pattern of BCC-structured W in [111] zone axis. (c-e) Measurement of the lattice distances of 5 atomic columns along <112> directions. The lattice distances of the two neighboring atomic columns on along <112> directions are measured to be 2.6 Å, and the interplaner angles between the edge-on {111} planes are determined to be 60°.

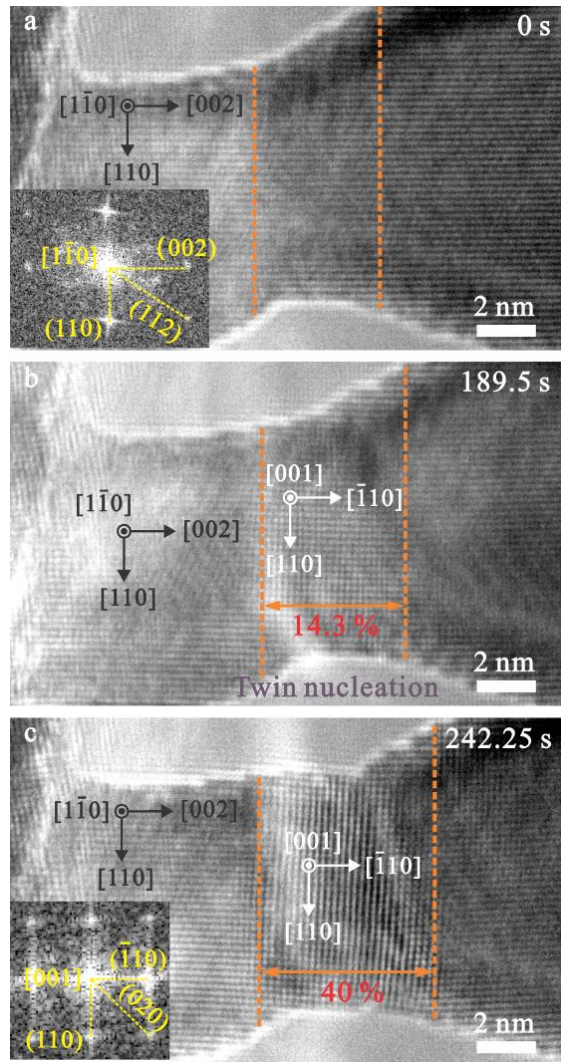

**Figure 7. Deformation twinning in W nanocrystal at room temperature.** (a) TEM image of the pristine 8.2-nm-diameter W nanocrystal under [002] tensile loading at the strain rate of $10^{-3}$ s$^{-1}$. The viewing direction is [1$\bar{1}$0]. The dashed lines indicate the gauge section used for calculating the elongation strain of the nanocrystal. The inset in (a) is the fast Fourier transform pattern of the pristine W nanocrystal. (b) Nucleation of a deformation twin resulting in the uniform elongation of 14.3%. (c) The deformed W nanocrystal exhibited an uniform elongation of nearlly 40% caused by deformation twinning. The inset in (c) is the fast Fourier transform pattern of the twinned W nanocrystal showing lattice reorientation from [002] to [$\bar{1}$10] direction.

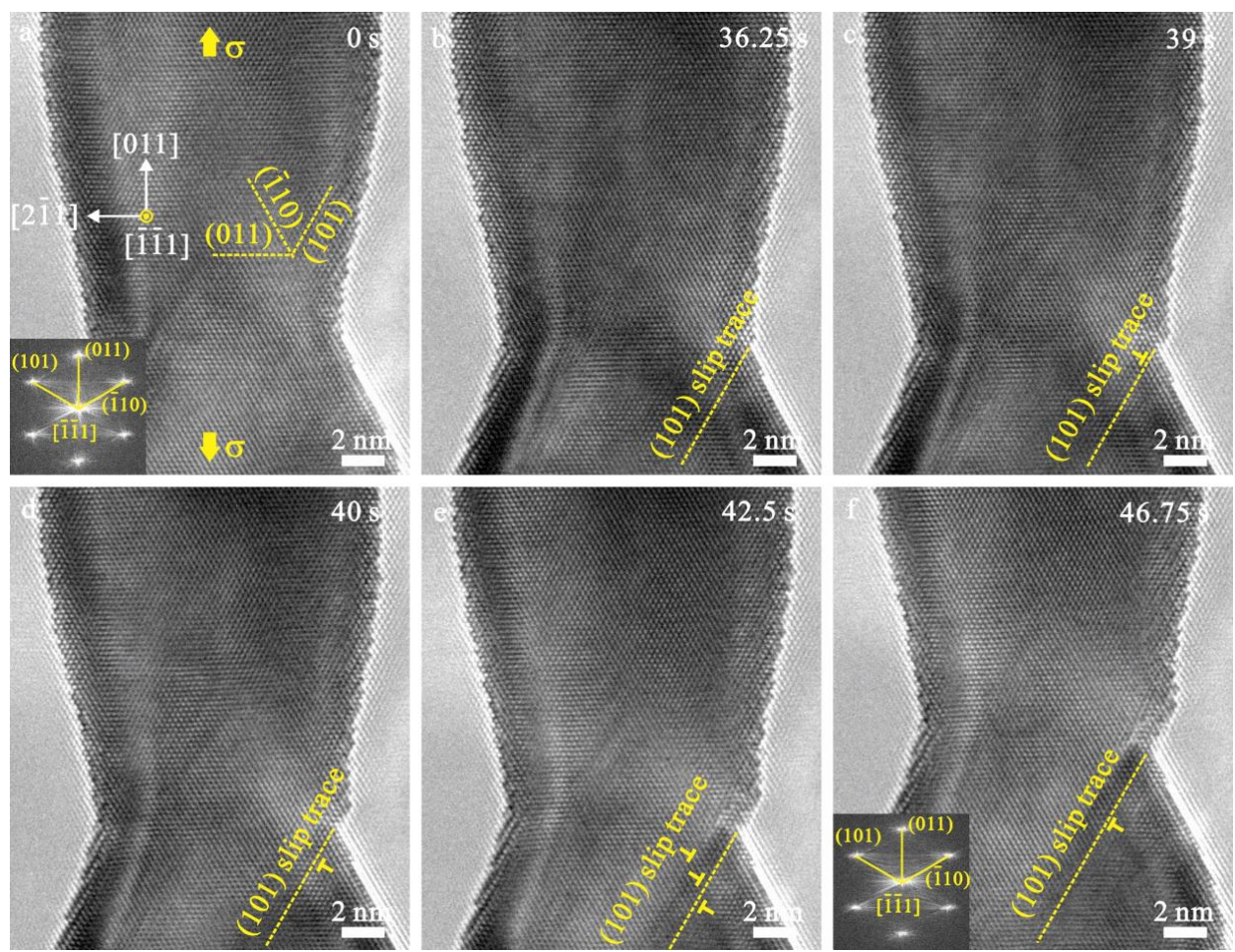

**Figure 8. Dislocation-mediated plasticity in a 12.1-nm-diameter W nanocrystal under <110> tensile loading at room temeperature. (a) TEM image of the pristine W nanocrystal loaded at the strain rate of $10^{-3}$ s$^{-1}$. The viewing direction is $[\bar{1}\bar{1}1]$ zone axis. The inset in (a) is the fast Fourier transform pattern of the pristine W nanocrystal. (b-f) A series of TEM images showing the occurrence of perfect dislocation slip on (101) planes in the W nanocrystal upon tensile loading. The inset in (f) is the fast Fourier transform patern of the deformed W nanocrystal demonstrating that no phase transformation occurs.**

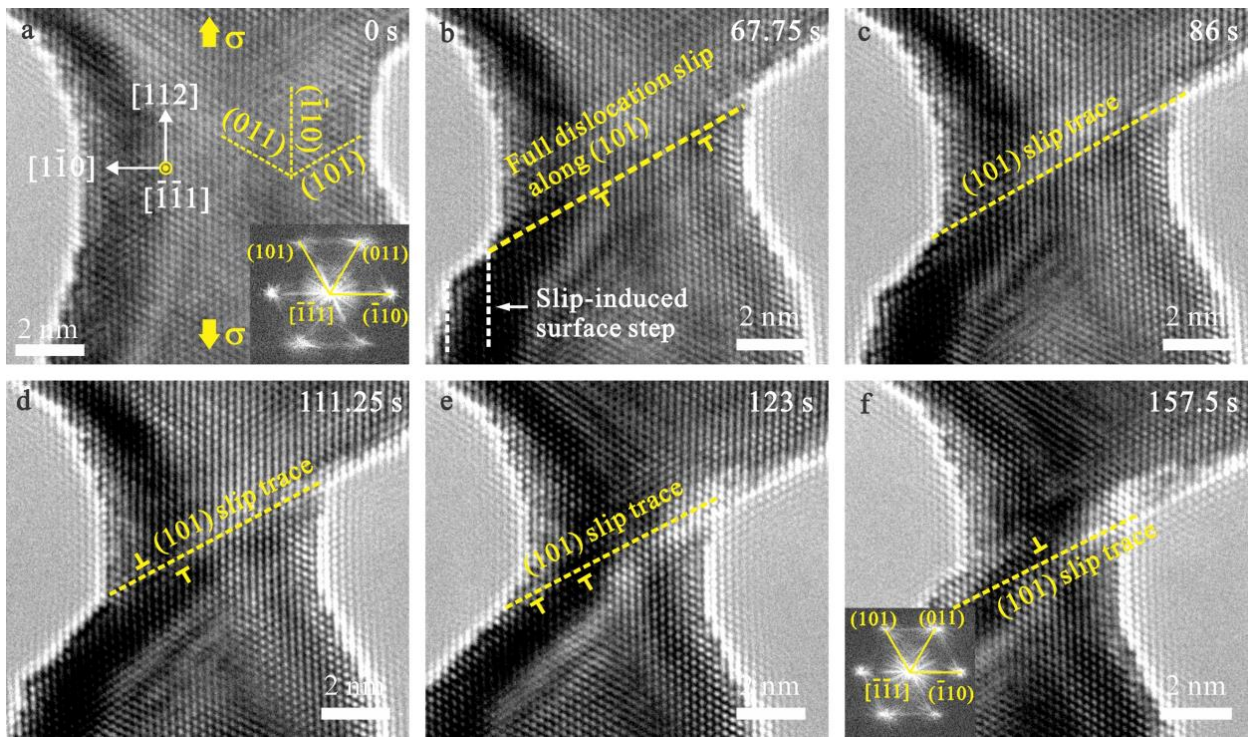

**Figure 9.** Dislocation-mediated plasticity in a 8.7-nm-diameter W nanocrystal under <112> tensile loading at room temeperature. (a) TEM image of the pristine W nanocrystal loaded at the strain rate of $10^{-3}$ s$^{-1}$. The viewing direction is along [$\bar{1}\bar{1}$1]. (b-f) Sequential TEM images showing that dislocation slip occurs in the W nanocrystal during tensile test. The insets in (a) and (f) are the fast Fourier transform images of the pristine and deformed W nanocrystals, repectively, demonstrating that no phase transformation occurs.

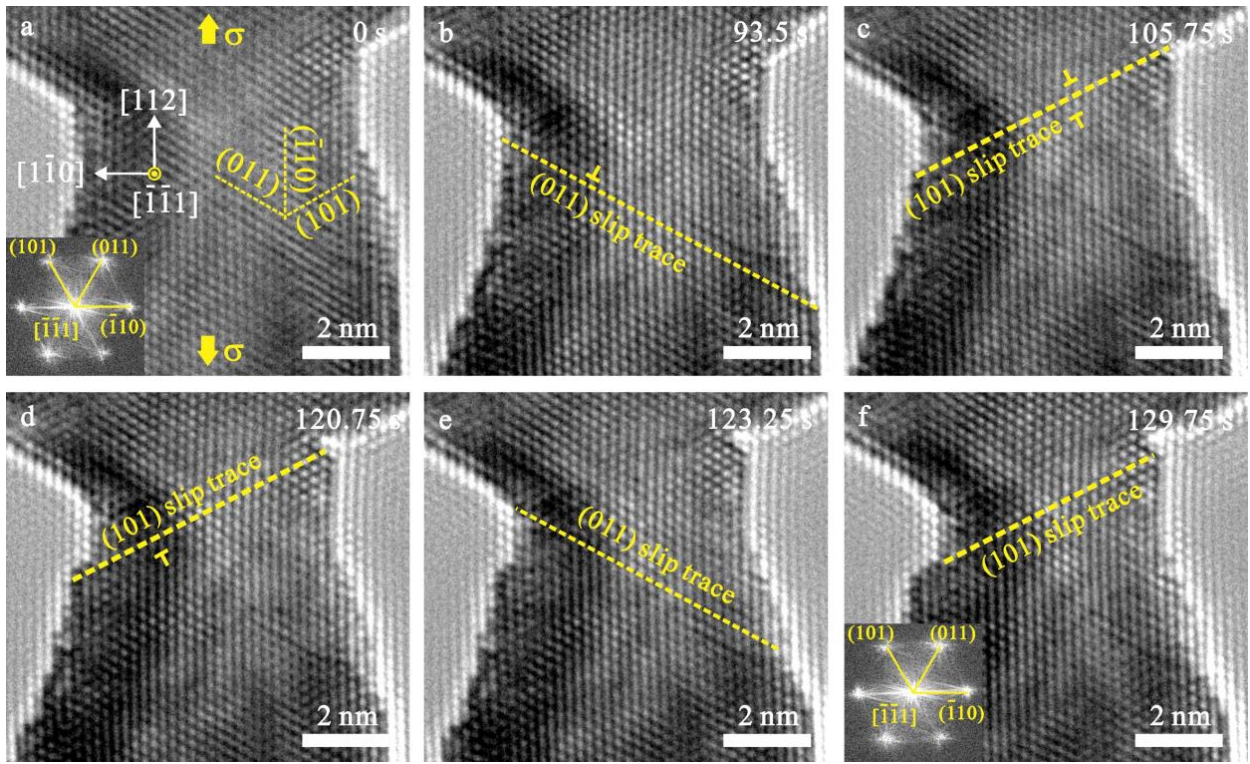

**Figure 10. Dislocation-mediated plasticity in a 6.7-nm-diameter W nanocrystal under [112] tensile loading within the temeperature range from 912 K to 915 K. (a) TEM image of the pristine W nanocrystal loaded at the strain rate of $10^{-3}$ s$^{-1}$. The viewing direction is along [$\bar{1}\bar{1}1$] zone axis. The inset in (a) is the fast Fourier transform pattern of the pristine W nanocrystal. (b-f) Sequential TEM images showing ththe occurrence of perfect dislocation slip on (101) and (011) planes in the W nanocrystal upon tensile deformation. The inset in (f) is the fast Fourier transform pattern of the deformed W nanocrystal demonstrating that no phase transformation occurs.**